\documentclass{iaus}
\usepackage{graphicx}
\def\spose#1{\hbox to 0pt{#1\hss}}
\def\lta{\mathrel{\spose{\lower 3pt\hbox{$\mathchar"218$}}\raise 2.0pt\hbox{$\mathchar"13C$}}}
\def\gta{\mathrel{\spose{\lower 3pt\hbox{$\mathchar"218$}}\raise 2.0pt\hbox{$\mathchar"13E$}}}
\def\etal{et~al.\ }
\def\kms{\,{\rm km\,s^{-1}}}
\newcommand{\HH}{H$_2$}
\newcommand{\msun}{\,{\rm M_\odot}}
\newcommand{\Tvir}{T_{\rm vir}}
\newcommand{\msub}{M_{\rm sub}}
\newcommand{\rvir}{r_{\rm vir}}


\def\beq{\begin{equation}}
\def\eeq{\end{equation}}
\def\HII{\hbox{H~$\scriptstyle\rm II\ $}}

\def\HH{H$_2$}
\def\Lya{Ly$\alpha\ $}
\def\vir{{\rm vir}}

\title[Formation and early evolution of MBHs] 
{Formation and early evolution of massive black holes}

\author[Piero Madau]   
{Piero Madau}

\affiliation{Department of Astronomy and Astrophysics, 
University of California, Santa Cruz, CA 95064\break email: 
pmadau@ucolick.org}

\pubyear{2006}
\volume{238}  
\pagerange{001--999}
\date{December 15, 2006}
\jname{Black Holes: from Stars to Galaxies -- across the Range of Masses}
\editors{V. Karas \& G. Matt, eds.}
\begin{document}

\maketitle

\begin{abstract}
The astrophysical processes that led to the formation of the first 
seed black holes and to their growth into the supermassive variety that
powers bright quasars at $z\sim 6$ are poorly understood.  
In standard $\Lambda$CDM hierarchical cosmologies, the earliest massive 
holes (MBHs) likely formed at redshift $z\gta 15$ at the 
centers of low-mass ($M\gta 5\times 10^5\msun$) dark matter ``minihalos'',
and produced hard radiation by accretion. FUV/X-ray photons from 
such ``miniquasars'' may have permeated the universe more uniformly 
than EUV radiation, reduced gas clumping, and changed the 
chemistry of primordial gas. The role of accreting seed black holes in 
determining the thermal and ionization state of the intergalactic medium 
depends on the amount of cold and dense gas that forms and gets retained 
in protogalaxies after the formation of the first stars. The highest resolution 
N-body simulation to date of Galactic substructure shows that subhalos 
below the atomic cooling mass were very inefficient at forming stars. 
\end{abstract}

\firstsection 

\section{Introduction}

The strong link observed between the masses of MBHs at the center of 
most galaxies and the gravitational potential wells
that host them (Ferrarese \& Merritt 2000; Gebhardt \etal 2000) suggests 
a fundamental mechanism for 
assembling black holes and forming spheroids in galaxy halos. In popular 
cold dark matter (CDM) hierarchical cosmologies, small-mass subgalactic 
systems form first to merge later into larger and larger structures.
According to these theories, some time beyond a redshift of 15 or so, the gas within 
halos with virial temperatures $T_\vir\gta 10^4\,$K -- or, equivalently, 
with masses $M\gta 10^8\, [(1+z)/10]^{-3/2}\,\msun$ -- cooled rapidly due
to the excitation of hydrogen \Lya and fragmented. Massive stars formed 
with some initial mass function (IMF) and synthesized heavy elements. 
Such early stellar systems, aided perhaps by a population of accreting 
black holes in their nuclei, generated the ultraviolet radiation and mechanical energy 
that reheated and reionized the cosmos. It is widely believed that 
collisional excitation of molecular hydrogen may have allowed gas in even 
smaller systems -- virial temperatures of a thousand K, corresponding to masses 
around $5\times 10^5\,[(1+z)/10]^{-3/2}\,\msun$ --  to cool and form stars 
at even earlier times (e.g. Tegmark \etal 1997). 

The first generation of seed MBHs must have formed in subgalactic units far up 
in the merger hierarchy, well before the bulk of the stars observed today: 
this is in order to have had sufficient time to build up via gas accretion 
a mass of several $\times 10^9\,\msun$ by $z=6.4$, the redshift of the most distant
quasars discovered in the {\it Sloan Digital Sky Survey} (SDSS) (Figure 1; Volonteri
\& Rees 2005). In hierarchical cosmologies, the ubiquity of MBHs in nearby luminous
galaxies can arise even if only a small fraction of halos harbor seed holes
at very high redshift (Menou, Haiman, \& Narayanan 2001).
The origin and nature of this seed population remain uncertain. Numerical simulations 
of the fragmentation of primordial molecular clouds in hierarchical 
cosmologies all show the formation of Jeans unstable clumps with masses 
exceeding a few hundred solar masses, with the implication that the 
resulting initial mass function is likely to be biased to very 
massive ``Population III'' stars (Abel, Bryan, \& Norman 2002; Bromm,
Coppi, \& Larson 2002). Barring any fine tuning of the IMF, 
intermediate-mass black holes -- 
with masses above the 4--18$\,\msun$ range of known stellar-mass 
holes  -- may then be the inevitable endproduct of the first episode 
of pregalactic star formation.


\begin{figure}
\includegraphics[width=0.5\textwidth]{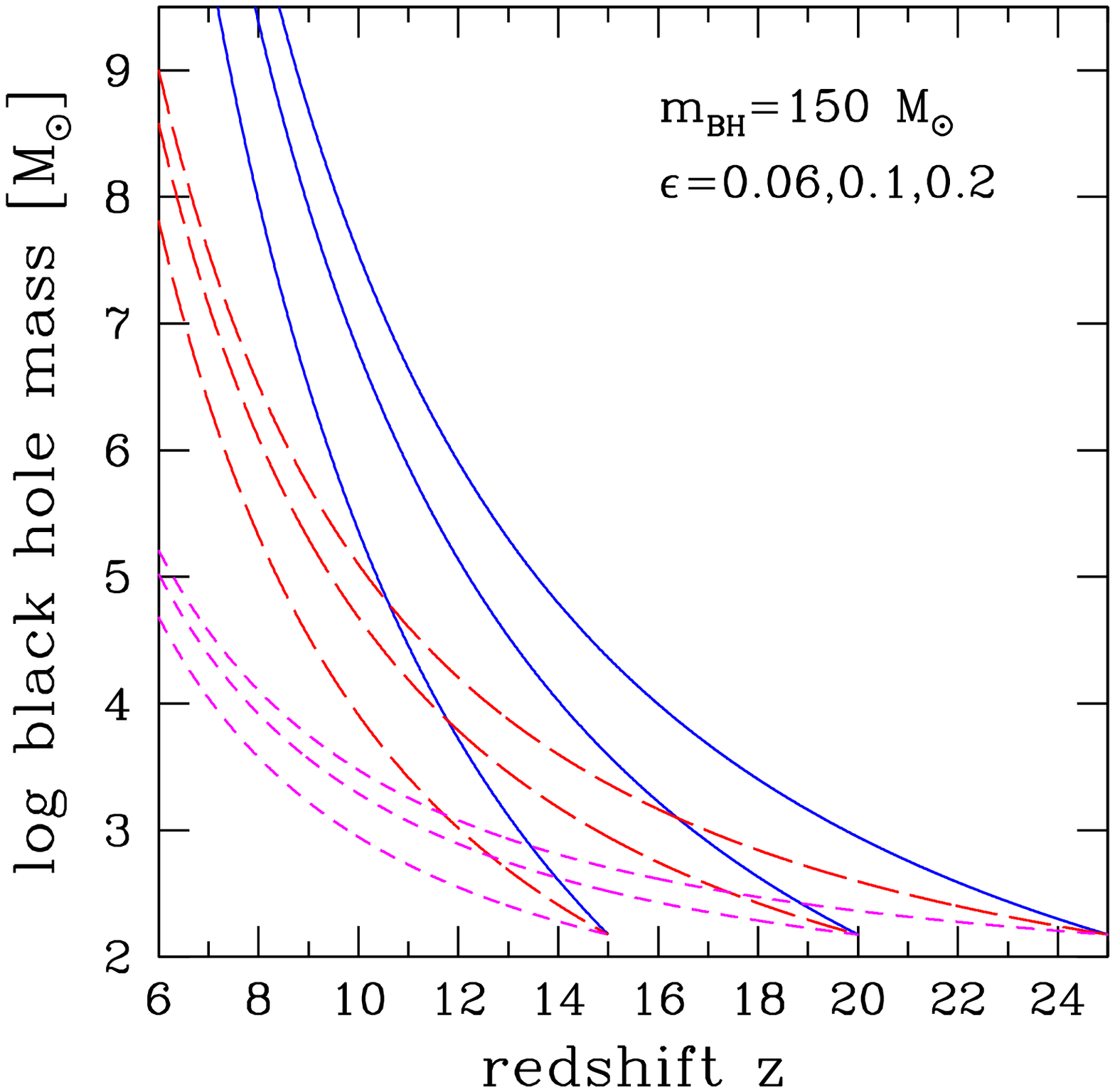}
\includegraphics[width=0.5\textwidth]{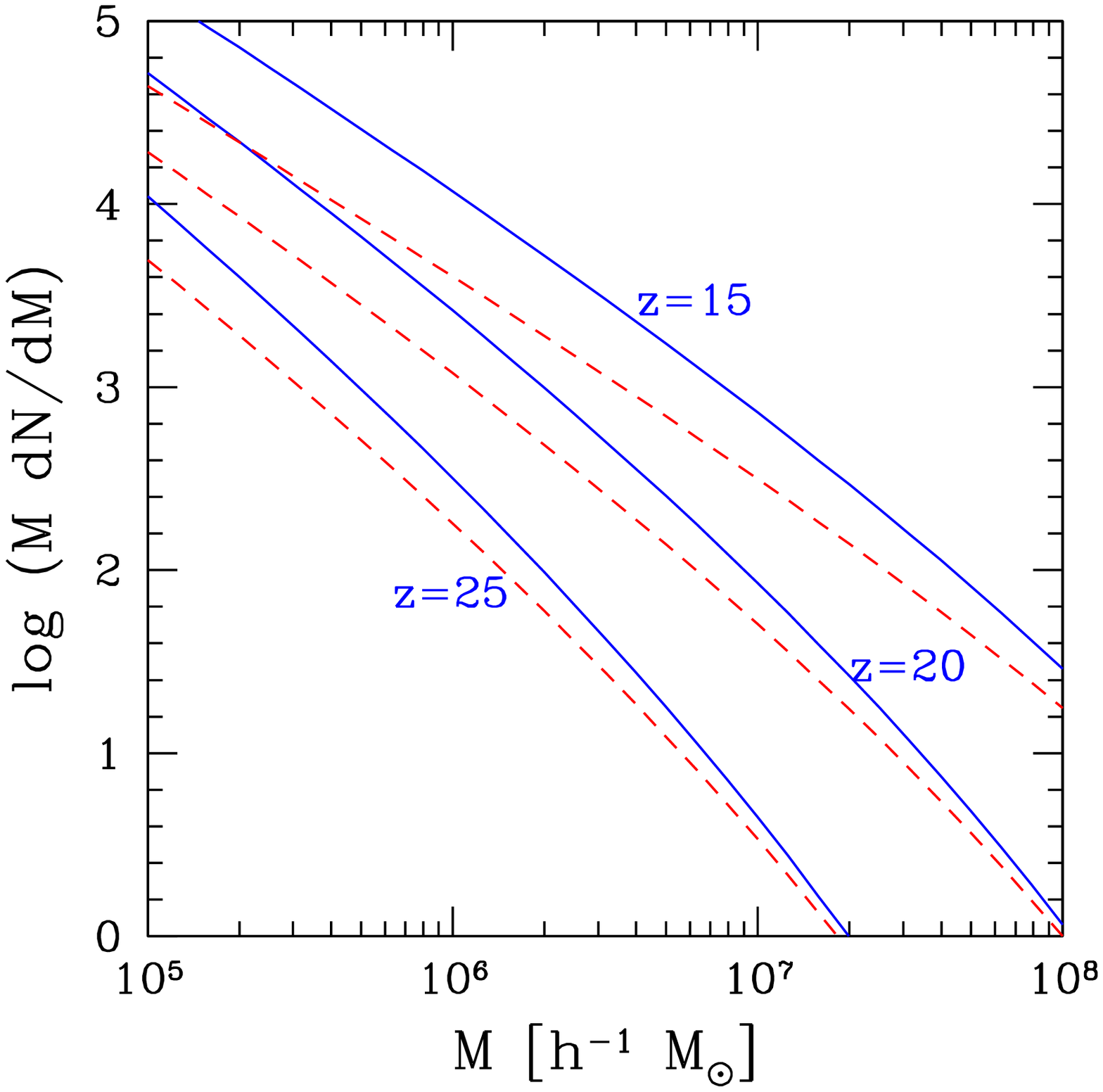}
\caption{{\it Left:} Growth of MBHs from early epochs down to $z=6$, the redshift
of the most distant SDSS quasars. The three sets of curves assume Eddington-limited accretion
with radiative efficiency $\epsilon=0.06$ ({\it solid lines}), 0.1 ({\it long-dashed lines}),
and 0.2 ({\it short-dashed lines}). Gas accretion starts at $z=15, 20, 25$ onto 
a seed black of mass $m_{\rm BH}=150\,\msun$.    
{\it Right:} Mass function of progenitor dark matter halos of mass $M$ at $z=15, 20, 25$ 
which, by the later time $z_0$, will have merged into a larger host of mass $M_0$. 
{\it Solid curves:} $z_0=0.8$,
$M_0=10^{12}\,h^{-1}\,\msun$ (``Milky Way'' halo). {\it Dashed curves:}
$z_0=3.5$, $M_0=2\times 10^{11}\,h^{-1}\,\msun$ (older ``bulge'').
If one seed hole formed in each $\sim 10^6\,\msun$ minihalo collapsing at
$z\sim 20$ (and triple hole interactions and binary coalescences were
neglected), several thousands relic seed black holes and their descendants would
be orbiting within the halos of present-day galaxies. (From Madau \& Rees 2001.)
}
\end{figure}

\section{MBHs as Population III remnants}

The first stars in the universe formed out of metal-free gas, in dark
matter ``minihalos'' collapsing from the high-$\sigma$ peaks of the primordial
density field at redshift $z>15$ or so. Gas condensation in the first baryonic 
objects was possible through the formation of H$_2$ molecules, which cool via 
roto-vibrational transitions down to temperatures of a couple hundred kelvins. 
At zero metallicity mass loss through radiatively-driven stellar
winds or nuclear-powered stellar pulsations is expected to be negligible, and
Pop III stars will likely die losing only a small fraction of their mass
(except for $100<m_*<140\,\msun$). Nonrotating
very massive stars in the mass window $140\lta m_*\lta 260\,\msun$ will disappear
as pair-instability supernovae (Bond, Arnett, \& Carr 1984), leaving no 
compact remnants and polluting the universe with the first heavy 
elements (Oh et al. 2001).\footnote{Since primordial metal enrichment was intrinsically a local process,
pair-instability SNe may occur in pockets of metal-free gas over a broad range of redshifts. 
The peak luminosities of typical pair-instability SNe are only slightly greater than
those of Type Ia, but they remain bright much longer ($\sim$ 1 year)
and have hydrogen lines (Scannapieco et al. 2005).}\
Stars with $40<m_*<140\,\msun$ and $m_*>260\,\msun$ are predicted instead
to collapse to black holes with masses exceeding half of the initial
stellar mass (Heger \& Woosley 2002).  
Since they form in high-$\sigma$ rare density peaks, relic seed black holes are expected
to cluster in the bulges of
present-day galaxies as they become incorporated through a series of
mergers into larger and larger systems (see Figure 1).
The presence of a small cluster of Population III holes in galaxy nuclei may have 
several interesting consequences associated with tidal captures of ordinary
stars (likely followed by disruption), capture by the central supermassive hole,
and gravitational wave radiation from such coalescences. Accreting pregalactic 
seed holes  may be detectable as ultra-luminous, off-nuclear X-ray sources (Madau
\& Rees 2001). 

\section{The first miniquasars}

Physical conditions in the central potential wells of young and gas-rich 
protogalaxies may have been propitious for black hole gas accretion.  
The presence of accreting black holes powering Eddington-limited 
miniquasars at such crucial formative stages in the evolution of the universe
may then present a challenge to models of the epoch of first light and of the
thermal and ionization early history of the intergalactic medium
(IGM), as reheating from the first stars and their remnants likely played a key
role in structuring the IGM and in regulating gas
cooling and star formation in pregalactic objects. Energetic photons
from miniquasars may make the low-density IGM warm and weakly ionized
prior to the epoch of reionization breakthrough (Madau \etal 2004;
Ricotti, Ostriker, \& Gnedin 2005). X-ray radiation may boost the
free-electron fraction and catalyze the formation of \HH\ molecules in
dense regions, counteracting their destruction by UV Lyman-Werner
radiation (Haiman, Abel, \& Rees 2000; Machacek, Bryan, \& Abel 2003).  Or it
may furnish an entropy floor to the entire IGM, preventing gas
contraction and therefore impeding rather than enhancing \HH\
formation (Oh \& Haiman 2003).  Photoionization heating may evaporate
back into the IGM some of the gas already incorporated into halos
with virial temperatures below a few thousand kelvins (Barkana \& Loeb
1999). The detailed consequences of all these effects is poorly understood.
In the absence of a UV photodissociating flux and of ionizing X-ray radiation, 
3D simulations show that the fraction of cold, 
dense gas available for star formation or accretion onto seed holes exceeds 
20\% for halos more massive than $10^6\,\msun$ (Figure 2; Machacek \etal 2003; 
Kuhlen \& Madau 2005). Since a zero-metallicity 
progenitor star in the range $40<m_*<500\,\msun$ emits about 80,000 photons above 1 ryd 
per stellar baryon (Schaerer 2002), the ensuing ionization front may overrun the 
host halo, photoevaporating most of the surrounding gas. Black hole remnants of the first 
stars that created \HII regions are then unlikely to accrete significant mass until new
cold material is made available through the hierarchical merging of
many gaseous subunits.

\begin{figure}
\begin{center}
\includegraphics[width=0.8\textwidth]{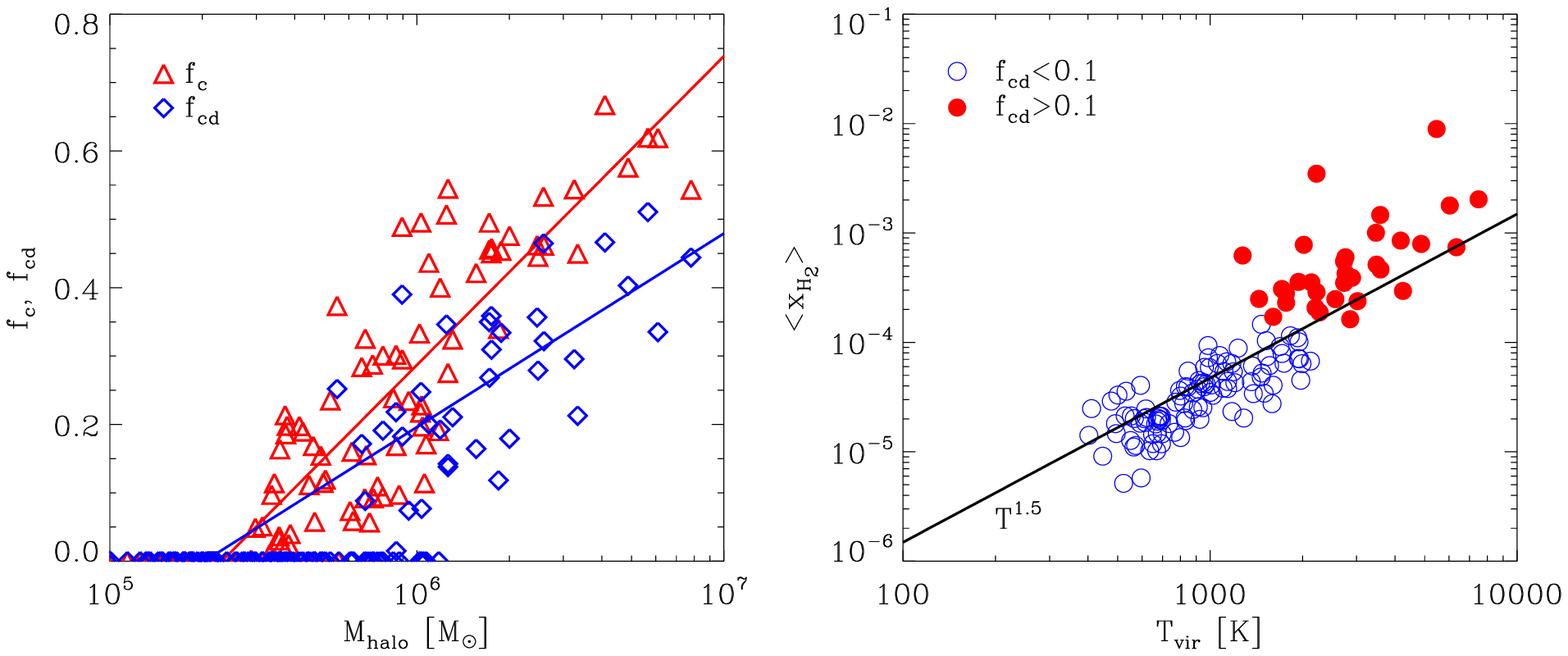}
\end{center}
\caption{{\it Left:} Fraction of cold and cold$+$dense gas within the virial
radius of all halos indentified at $z=17.5$ with $\Tvir>400$K, as a
function of halo mass. {\it Triangles:} $f_c$, fraction of halo gas
with $T<0.5\,\Tvir$ and $\delta>1000$ that cooled via roto-vibrational
transitions of \HH.  {\it Diamonds:} $f_{\rm cd}$, fraction of gas
with $T<0.5\,\Tvir$ and $\rho> 10^{19}\,\msun$ Mpc$^{-3}$ that is
available for star formation. The straight lines represent mean
regression analyses of $f_c$ and $f_{\rm cd}$ with the logarithm of
halo mass. {\it Right:} Mass-weighted mean \HH\ fraction as a function
of virial temperature for all halos at $z=17.5$ with $\Tvir>400\,$K
and $f_{\rm cd}<0.1$ ({\it empty circles}) or $f_{\rm cd}>0.1$ ({\it
filled circles}). The straight line marks the scaling of the
temperature-dependent asymptotic molecular fraction. (From Kuhlen \& Madau 2005.)
}
\end{figure}

High-resolution hydrodynamics simulations
of early structure formation in $\Lambda$CDM cosmologies are a powerful tool to
track in detail the thermal and ionization history of a clumpy IGM and guide studies
of early reheating. In Kuhlen \& Madau (2005) we used a modified version of ENZO, a grid-based 
hybrid (hydro$+$N-body) code developed by Bryan \& Norman
(see http://cosmos.ucsd.edu/enzo/) to solve the cosmological hydrodynamics equations
and simulate the effect of a miniquasar turning on at very high redshift. 
The simulation follows the non-equilibrium chemistry of the
dominant nine species (H, H$^+$, H$^-$, e, He, He$^+$, He$^{++}$, H$_2$, and H$_2^+$)
in primordial gas, and includes radiative losses from atomic and molecular line
cooling. At $z=21$, a miniquasar powered by a 150 $\msun$ black hole accreting
at the Eddington rate is turned on in the protogalactic halo. The miniquasar
shines for a Salpeter time (i.e. down to $z=17.5$) and is a copious source
of soft X-ray photons,
which permeate the IGM more uniformly than possible with extreme ultraviolet (EUV,
$\ge 13.6\,$eV) radiation. After one Salpeter
timescale, the miniquasar has heated up the simulation box to a
volume-averaged temperature of 2800 K. The mean electron and \HH\
fractions are now 0.03 and $4\times10^{-5}$: the latter is 20 times
larger than the primordial value, and will delay the buildup of a
uniform UV photodissociating background. The net effect of the X-rays
is to reduce gas clumping in the IGM by as much as a factor of
3. While the suppression of baryonic infall and the photoevaporation
of some halo gas lower the gas mass fraction at overdensities $\delta$
in the range 20-2000, enhanced molecular cooling increases the amount
of dense material at $\delta>2000$. In many halos within the
proximity of our miniquasar the \HH-boosting effect of X-rays is too
weak to overcome heating, and the cold and dense gas mass actually
decreases. There is little evidence for an entropy floor preventing
gas contraction and \HH\ formation: instead, molecular
cooling can affect the dynamics of baryonic material before it has
fallen into the potential well of dark matter halos and virialized.
Overall, the radiative feedback from X-rays enhances gas cooling in
lower-$\sigma$ peaks that are far away from the initial site of star
formation, thus decreasing the clustering bias of the early
pregalactic population, but does not appear to dramatically reverse or
promote the collapse of pregalactic clouds as a whole.

\begin{figure}
\begin{center}
\includegraphics[width=0.7\textwidth,angle=90]{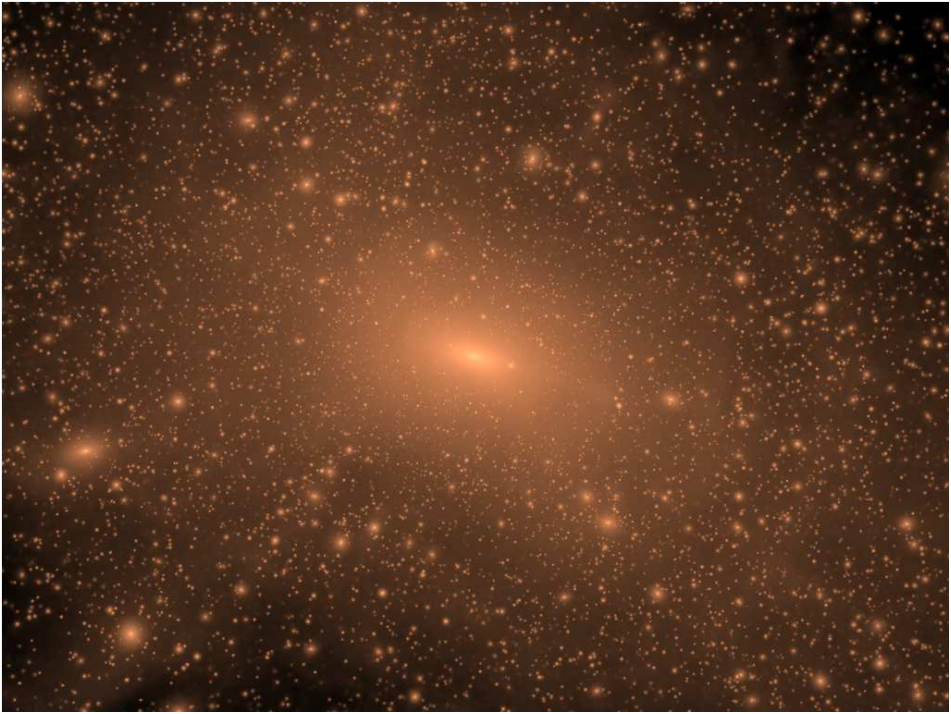}
\end{center}
\caption{Projected dark matter density-squared map of our simulated Milky Way-size 
halo (``Via Lactea'') at the present epoch. The image covers
an area of 800 $\times$ 600 kpc, and the projection goes through a 600
kpc-deep cuboid containing a total of 110 million particles. The
logarithmic color scale covers 20 decades in density-square. (From 
Diemand, Kuhlen, \& Madau 2006.)
}
\end{figure}

\section{Near-field cosmology and dark satellites}

Despite much recent progress in our understanding of the formation of  early 
cosmic structure and the high-redshift universe, many fundamental
questions related to the astrochemistry of primordial gas and the formation
and evolution of halo+MBH systems remain only partially answered. 
N-body$+$hydrodynamical simulations  are unable to predict, for example,
the efficiency with which the first gravitationally collapsed objects lit up the
universe at cosmic dawn, and treat the effects of the energy input
from the earliest generations of sources on later ones only in an 
approximate way.  The {\it Wilkinson Microwave Anisotropy Probe} 3-year data require 
the universe to be fully reionized by redshift $z_{\rm ion}=11\pm 2.5$
(Spergel \etal 2006), an indication
that significant star-formation activity started at early cosmic times.
We infer that Population III massive stars  and perhaps miniquasars must have been shining when 
the universe was less than 350 Myr old, but we remain uncertain about the nature of their host
galaxies and the impact they had on their 
environment and on the formation and structure of more massive systems.

\begin{figure}
\includegraphics[width=0.5\textwidth]{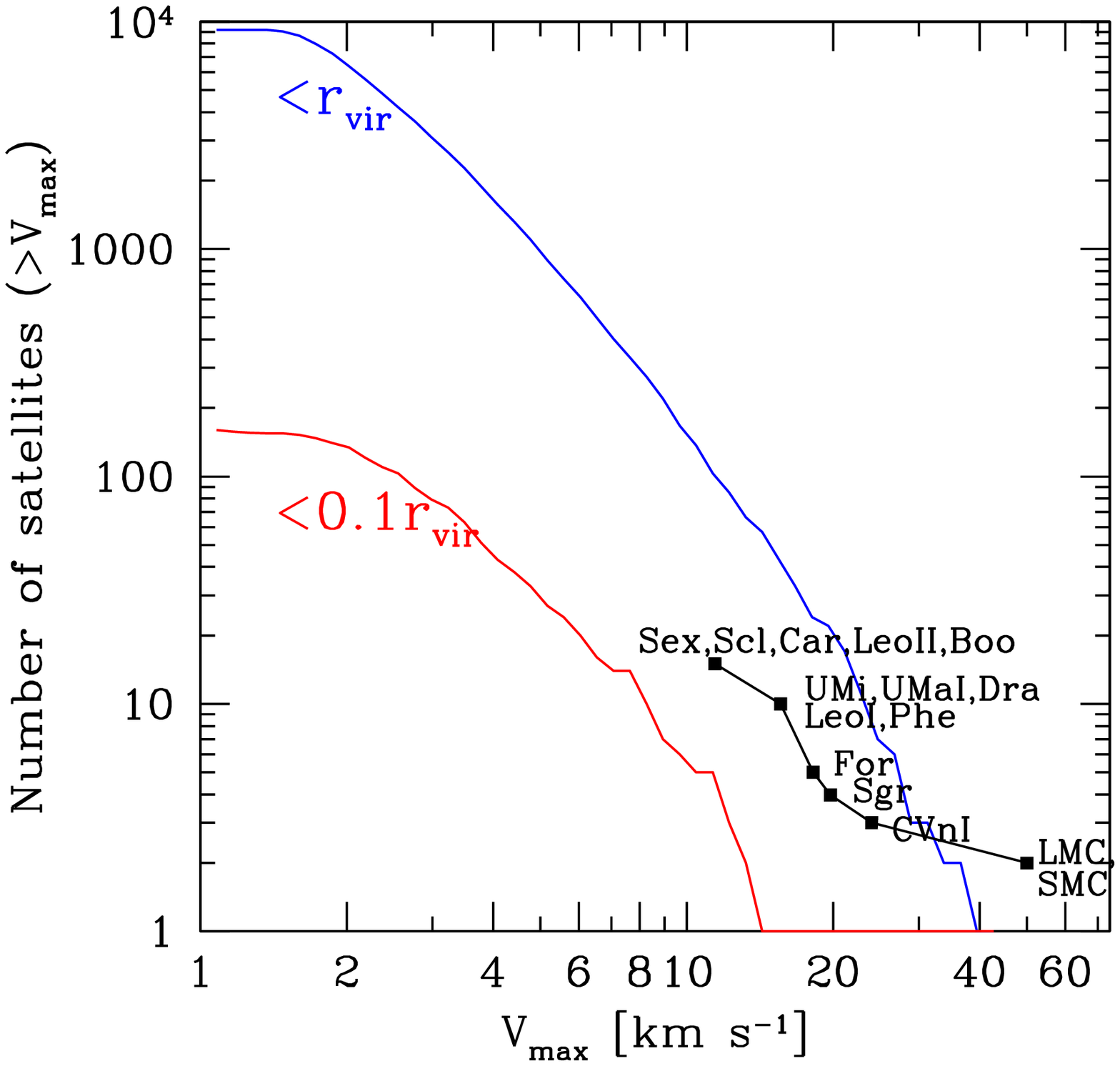}
\includegraphics[width=0.5\textwidth]{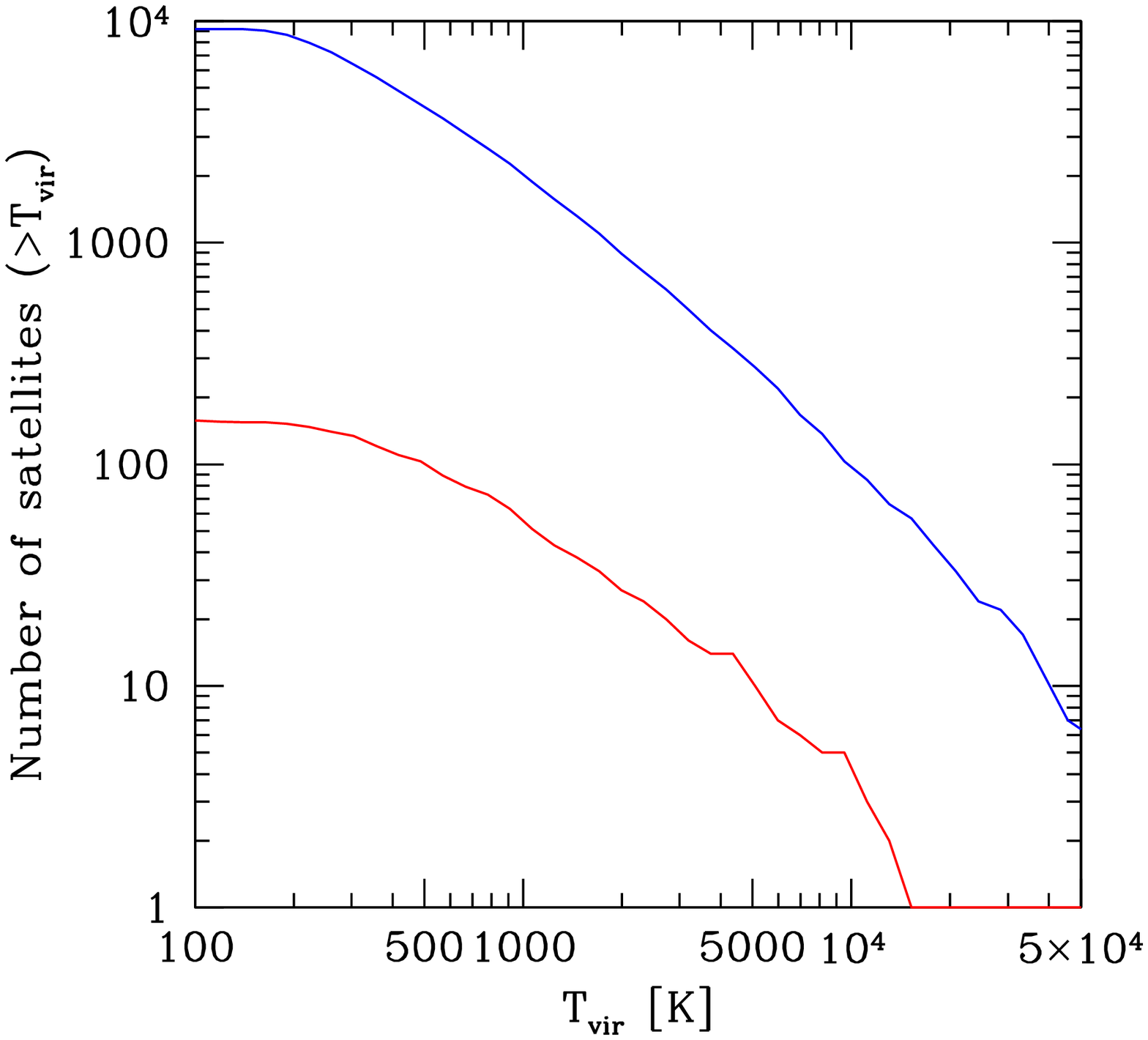}
\caption{{\it Left:} Cumulative peak circular velocity function for all subhalos within 
Via Lactea's $\rvir$ ({\it upper curve}) and for the subpopulation within 
the inner $0.1\,\rvir$ ({\it lower curve}). {\it Solid line with points:} observed 
number of dwarf galaxy satellites around the Milky Way. {\it Right:} Same plotted versus 
virial temperature $T_{\rm vir}$.
}
\end{figure}

While the above cosmological puzzles can be tackled directly by studying distant objects,
it has recently become clear that many of today's ``observables'' within the Milky Way 
and nearby galaxies relate to events occurring at very high redshifts, during and soon 
after the epoch of 
reionization (see Moore et al. 2006 and references therein). In this sense, galaxies in the 
Local Group can provide a crucial diagnostic link to the physical processes that govern 
structure formation and evolution in the early universe, an approach termed ``near-field
cosmology". It is now well established that the hierarchical mergers that form
the halos surrounding galaxies are rather inefficient, leaving 
substantial amounts of stripped halo cores or ``subhalos'' orbiting
within these systems. Small halos
collapse at high redshift when the universe is very dense, so their
central densities are correspondingly high. When these merge
into larger hosts, their high densities allow them to resist the
strong tidal forces that acts to destroy them. Gravitational
interactions appear to unbind most of the mass associated with 
the merged progenitors, but a significant fraction of these small 
halos survives as distinct substructure. 

In order to recognize the basic building blocks of galaxies in the
near-field, we have recently completed ``Via Lactea'', the most detailed simulation 
of Milky Way CDM substructure to date (Diemand, Kuhlen, \& Madau 2006). 
The simulation resolves a Milky Way-size halo with 85 million particles 
within its virial radius $\rvir$, and was run for $320,000$ cpu-hours on NASA's 
SGI Altix supercomputer \textit{Columbia}.
Figure 3 shows a projected dark matter density squared map of a $800\times600$ 
kpc region of this simulation at the current epoch. The high resolution region was 
sampled with 234 million particles of mass $2.1\times 10^4\msun$, evolved
with a force resolution of 90 pc, and centered on an isolated halo that 
had no major merger after $z=1.7$, making it a suitable host for a Milky Way-like
disk galaxy. 

Approximately 10,000 surviving satellite halos can be identified at the present epoch:
this is more than an order of magnitude larger than found in previous $\Lambda$CDM
simulations. Their cumulative mass function is well-fit by $N(>\msub)\propto 
\msub^{-1}$ down to $\msub=4\times 10^6\,\msun$. Sub-substructure
is apparent in all the larger satellites, and a few dark matter lumps are now resolved
even in the solar vicinity. In Via Lactea, the number of dark satellites
with peak circular velocities above $5\,\kms$ ($10\,\kms$) exceeds 800 (120).  As shown
in Figure 4, such finding appears to exacerbate the so-called ``missing satellite problem'', the 
large mismatch between the twenty or so dwarf satellite galaxies observed 
around the Milky Way and the predicted large number of CDM subhalos (Moore et al.
1999; Klypin et al. 1999).  Solutions involving feedback mechanisms that make halo 
substructure very inefficient in forming stars offer a possible way out (e.g. Bullock, 
Kravtsov, \& Weinberg 2000; Kravtsov, Gnedin, \& Klypin 2004; Moore et al. 2006). In this 
case seed black holes may not have grown efficiently in small minihalos just 
above the cosmological Jeans mass, and gas accretion may have had to await the buildup 
of more massive galaxies (with virial temperatures above the threshold for atomic 
cooling). 

\section{MBH spins}

As discussed in the Introduction, growing very large
MBHs at high redshift from small seeds  requires low accretion efficiencies,
or, equivalently, modest black hole spins.
The spin of a MBH is also expected to have implications for the
direction of jets in active nuclei and to determine the
innermost flow pattern of gas accreting onto Kerr holes (Bardeen \&
Petterson 1975). The coalescence of two spinning black holes in a radio
galaxy may cause a sudden reorientation of the jet direction, perhaps
leading to the so-called ``winged'' or ``X-type'' radio sources
(Merritt \& Ekers 2002).  MBH spins are determined by the competition between a number of
physical processes.  Black holes forming from the gravitational
collapse of very massive stars endowed with rotation will in general
be born with non-zero spin (e.g. Fryer, Woosley, \& Heger 2002).  An
initially non-rotating hole that increases its mass by (say) 50\% by
swallowing material from an accretion disk may be spun up to $a/m_{\rm
BH}=0.84$. While the coalescence of two non-spinning
black holes of comparable mass will immediately drive the spin
parameter of the merged hole to $a/m_{\rm BH}\gta 0.8$ (e.g. Gammie,
Shapiro, \& McKinney 2004), the capture of smaller companions in
randomly-oriented orbits may spin down a Kerr hole instead (Hughes \&
Blandford 2003).

\begin{figure}
\begin{center}
\includegraphics[width=0.6\textwidth]{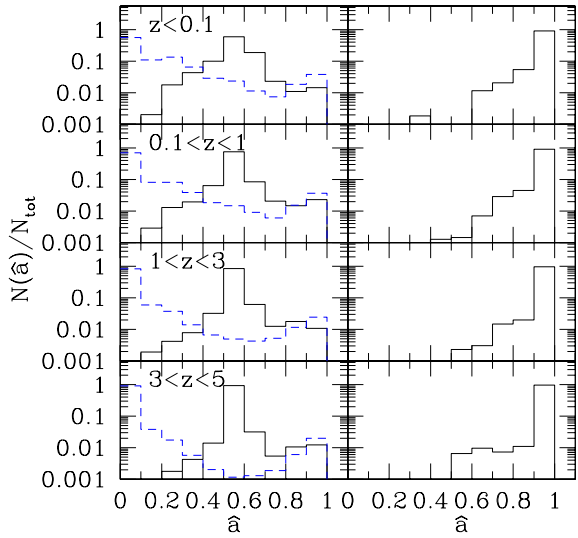}
\end{center}
\caption{\footnotesize Distribution of MBH spins in different redshift intervals.
{\it Left panel:} effect of black hole binary coalescences only.
{\it Solid histogram:} seed holes are born with $\hat a\equiv a/m_{\rm BH}=0.6$.
{\it Dashed histogram:} seed holes are born non-spinning.
{\it Right panel:} spin distribution from binary coalescences and gas accretion.
Seed holes are born with $\hat a=0.6$, and are efficiently spun up
by accretion via a thin disk.
}
\end{figure}

In  Volonteri et al. (2004) we made an attempt at estimating 
the distribution of MBH spins and its evolution with cosmic time in the context of
hierarchical structure formation theories, following the combined effects of black
hole-black hole coalescences and accretion from a gaseous thin disk on the
magnitude and orientation of MBH spins. 
Binary coalescences appear to cause no significant systematic
spin-up or spin-down of MBHs: because of the relatively flat
distribution of MBH binary mass ratios in hierarchical models, 
the holes random-walk around the spin parameter they are endowed with at birth, and 
the spin distribution retains significant memory of the initial rotation of ``seed'' holes.
It is accretion, not binary coalescences, that dominates the spin
evolution of MBHs (Fig. 5). Accretion can lead to efficient spin-up of MBHs
even if the angular momentum of the inflowing material varies in
time, provided the fractional change of mass during each accretion episode
of a growing black hole is large. In this case, for a thin accretion disk, the hole is aligned
with the outer disk on a timescale that is much shorter than the
Salpeter time (Natarajan \& Pringle 1998), leading
to accretion via prograde equatorial orbits: most of the
mass accreted by the hole acts to spin it up, even if the orientation
of the spin axis changes in time. Under the combined effects of accretion and binary 
coalescences, we found that the spin distribution is heavily skewed towards
fast-rotating Kerr holes, is already in place at early epochs, and
does not change significantly below redshift 5. One way to avoid rapid rotation 
and produce slowly rotating fast-growing holes is
to assume ``chaotic feeding'' in which small amounts of material, with
$\Delta m\ll m_{\rm BH}$, are swallowed by the hole in successive
accretion episodes with random orientations (e.g., Moderski and Sikora
1996; Volonteri et al. 2004; King \& Pringle 2006). 

\bigskip\bigskip

I would like to thank all my collaborators for their contributions to the ideas
presented here. Support for this work was provided by NASA grant 
NNG04GK85G.

\end{document}